\documentclass[%
twocolumn,
superscriptaddress,
nofootinbib,
 amsmath,amssymb,
 aps,
 pra,
]{revtex4-2}

\usepackage{graphicx,subfigure}
\usepackage{dcolumn}
\usepackage{bm}
\usepackage[T1]{fontenc}
\usepackage[section]{placeins}
\RequirePackage{mathpazo} 
\usepackage{textcomp}
\usepackage{float}
\usepackage{braket}
\usepackage{amsmath}
\usepackage{mathtools}
\usepackage[normalem]{ulem}

\usepackage{hyperref}

\usepackage{xcolor}
\definecolor{red}{rgb}{0.9,0,0}
\definecolor{magenta}{rgb}{1.0,0,1.0}

\def\ak#1{\textbf{\color{orange}[#1]}}
\def\pj#1{\textbf{\color{magenta}[#1]}}

\usepackage{enumitem,amssymb}
\newlist{todolist}{itemize}{2}
\setlist[todolist]{label=$\square$}
\usepackage{pifont}

\usepackage{makecell}

\begin{document}

\title{Reducing the cost of energy estimation in the variational quantum eigensolver algorithm with robust amplitude estimation}

\author{Peter D. Johnson}
\affiliation{Zapata Computing, Inc., 100 Federal St., Boston, MA 02110, USA}
\author{Alexander A. Kunitsa}
\affiliation{Zapata Computing, Inc., 100 Federal St., Boston, MA 02110, USA}
\author{J\'er\^ome F. Gonthier}
\affiliation{Zapata Computing, Inc., 100 Federal St., Boston, MA 02110, USA}
\author{Maxwell D. Radin}
\affiliation{Zapata Computing, Inc., 100 Federal St., Boston, MA 02110, USA}
\author{Corneliu Buda}
\affiliation{Applied Chemistry and Physics Centre of Expertise, BP Group Research, 150 West Warrenville Road, Naperville, IL 60563, USA}
\author{Eric J. Doskocil}
\affiliation{Applied Chemistry and Physics Centre of Expertise, BP Group Research, 150 West Warrenville Road, Naperville, IL 60563, USA}
\author{Clena M. Abuan}
\affiliation{Digital Science and Engineering, BP Innovation and Engineering, 501 Westlake Park Blvd, Houston, TX 77079, USA}
\author{Jhonathan Romero}
\affiliation{Zapata Computing, Inc., 100 Federal St., Boston, MA 02110, USA}

\date {\today}

\begin{abstract}

Quantum chemistry and materials is one of the most promising applications of quantum computing.
Yet much work is still to be done in matching industry-relevant problems in these areas with quantum algorithms that can solve them.
Most previous efforts have carried out resource estimations for quantum algorithms run on large-scale fault-tolerant architectures, which include the quantum phase estimation algorithm.
In contrast, few have assessed the performance of near-term quantum algorithms, which include the variational quantum eigensolver (VQE) algorithm.
Recently, a large-scale benchmark study \cite{gonthier2020identifying} found evidence that the performance of the variational quantum eigensolver for a set of industry-relevant molecules may be too inefficient to be of practical use.
This motivates the need for developing and assessing methods that improve the efficiency of VQE. 
In this work, we predict the runtime of the energy estimation subroutine of VQE when using robust amplitude estimation (RAE) to estimate Pauli expectation values.
Under conservative assumptions, our resource estimation predicts that RAE can reduce the runtime over the standard estimation method in VQE by one to two orders of magnitude.
Despite this improvement, we find that the runtimes are still too large to be practical.
These findings motivate two complementary efforts towards quantum advantage: 1) the investigation of more efficient near-term methods for ground state energy estimation and 2) the development of problem instances that are of industrial value and classically challenging, but better suited to quantum computation.
\end{abstract}

\maketitle

\section{Introduction}

Identifying a promising candidate for practical quantum advantage lies at the frontier of modern quantum computing research \cite{reactions_on_QC, gonthier2020identifying, o2021efficient, alcazar2021quantum, stamatopoulos2021towards, stanisic2021observing, goings2022reliably}. Rapid improvement in quantum hardware \cite{Arute2019, kandala2019error, egan2021fault, abobeih2021fault, ryan2021realization} has given us hope that in the near future we will enter a new technological era marked by the widespread application of quantum algorithms to simulating chemistry and materials, solving differential equations, and modeling financial markets. Is it possible to predict the onset of quantum advantage for a particular industrially relevant problem? What are the quantum resources needed to achieve this using imperfect near-term devices? Our work is motivated by these questions in the context of chemistry simulation, focusing on the prototypical task of ground state energy estimation using the variational quantum eigensolver (VQE) algorithm \cite{Peruzzo2014}. 
The molecular ground state energy is useful for the important molecular modelling task of predicting the enthalpies of hydrocarbon combustion reactions. 
The starting point of this work is a related study of Gonthier et al.~\cite{gonthier2020identifying}, which raised the question: \emph{how do recent advances in quantum estimation \cite{wang2021minimizing} improve the pessimistic findings of VQE performance?}
Our work aims to answer this question.
We approach this problem from a resource estimation perspective, using the same set of molecules as in \cite{gonthier2020identifying} to facilitate comparison.

Ground state energy estimation is a fundamental problem in molecular quantum chemistry holding the key to multiple industrial applications, such as material design and accelerated drug discovery. It was among the first applications of both fault-tolerant~\cite{aspuru2005simulated} and near-term variational quantum algorithms~\cite{Peruzzo2014}. Several authors identified it as a promising candidate for industrially relevant quantum advantage, even though a specific problem for which it can be established remains a subject of debate ~\cite{Wecker2015,Burg2020,industrial_Q_advantage,Cade2020}. 

An important step along the path to quantum advantage for quantum chemistry and materials is the development of viable problem instances: identify industrially-relevant quantum systems for which solving the ground state energy problem with sufficient accuracy is classically intractable. Typical examples include so-called multireference systems, where the ground state cannot be even qualitatively described by the Hartree-Fock mean-field approximation. Motivated by this, previous work in resource estimation explored strongly correlated systems such as metalloenzyme co-factors~\cite{Burg2020, reactions_on_QC, goings2022reliably}, transition metal compounds~\cite{industrial_Q_advantage}, and the 2-D Fermi-Hubbard model~\cite{Cade2020}. In the present work, we instead look at systems that are well-described by single-reference methods. However, reaching sufficient accuracy can still be quite costly \cite{gonthier2020identifying}, and these systems have the advantage of readily available, accurate experimental data to gauge the accuracy of classical algorithms. These properties make these molecules a good benchmark set.

Most of these previous resource estimates in quantum chemistry have been devoted to assessing algorithms for large-scale fault-tolerant quantum computers.
This leaves open the question: how might near-term quantum computers be used to realize quantum advantage in quantum chemistry?


In the past decade, a host of quantum algorithms have been developed that are suited to the limitations of noisy intermediate-scale quantum (NISQ) devices.
The archetype of these methods is the variational quantum eigensolver (VQE) algorithm \cite{Peruzzo2014}.
VQE is a heuristic algorithm that leverages the variational principle of quantum mechanics to find the best approximation for a ground state of a given molecular Hamiltonian $H$ for a particular choice of a circuit ansatz $A$~\cite{Peruzzo2014,Kandala2017,Romero2018}.
The progress of these near-term quantum algorithms suggests an alternative route to discovering quantum advantage: begin with the capabilities of current quantum devices and determine what minimal improvements are needed for them to solve useful problems.
However, carrying out resource estimations for near-term quantum algorithms like VQE is challenging because they are heuristic:
unlike traditional quantum algorithms for the ground state energy estimation, such as QPE, VQE does not provide theoretical performance guarantees and needs to be benchmarked on a per case basis, taking into account the target precision and typical problem size. 

Recent work \cite{gonthier2020identifying} carried out a large-scale benchmark study on the resources needed to run VQE.
The authors considered a set of molecular Hamiltonians representing industrially-relevant hydrocarbon molecules.
They found that the runtime required to reach chemical accuracy (i.e. 1 kcal/mol) for the reaction energies is prohibitive under realistic assumptions for quantum gate times. 
This large runtime was mostly due to the time needed for the subroutine of energy expectation value estimation with standard sampling; the statistical nature of the energy estimation entails an inverse quadratic scaling of the number of measurements with respect to the target precision.

This statistical phenomenon is responsible for the so-called ``measurement problem'' of VQE with standard sampling: the number of statistical samples required to obtain sufficiently accurate energy estimates is large, leading to prohibitively large runtimes for the VQE algorithm. 
From the pessimistic findings regarding the measurement problem in \cite{gonthier2020identifying}, the authors concluded that techniques for speeding up the estimation subroutine used in VQE would be necessary in order to make the algorithm competitive with state-of-the-art classical methods on industry-relevant problem instances.
They suggest that techniques like \emph{Robust Amplitude Estimation} (RAE)~\cite{wang2021minimizing,ELF}, which increase the rate of information gain in estimation, will be needed to realize quantum advantage for the problem of ground state energy estimation.

RAE offers a new feature among near-term quantum algorithms for estimation: improvements in the quantum computer (as measured by reduction in gate error rates) translate into a proportional improvement in estimation performance (as measured by reduction in runtime).
Key to this feature is the robustness of the algorithm: it accommodates a degree of error in the operations by learning a model of the error's effect.

Recent work has investigated the use of similar techniques for the application of Monte Carlo integration in finance
\cite{alcazar2021quantum, giurgica2021low}.
To our knowledge, ours is the first effort to assess these methods for the application of quantum chemistry.
The objective of this work is to carry out a resource estimation for robust amplitude estimation applied to VQE energy estimation for the problem instances defined in \cite{gonthier2020identifying}.
Our resource estimates predict that, for the molecules considered, RAE gives between a 13 and 64 fold reduction in runtime over VQE.

The paper is structured as follows. In Section \ref{sec:background} we review the expectation value estimation techniques of standard sampling (as traditionally used in VQE) and robust amplitude estimation.
In Section \ref{sec:methods} we describe the methods used to carry out the resource estimations including algorithm performance modeling, circuit compilation, and the accounting of error correction overhead.
In Section \ref{sec:results} we describe the results of validating the RAE algorithm performance model and the results comparing the performance of standard sampling to robust amplitude estimation for the benchmark set of molecules.
Finally, we conclude in Section \ref{sec:outlook} with an outlook on future directions for discovering quantum advantage in quantum chemistry.

\section{Technical Background}
\label{sec:background}
\subsection{Standard sampling}
Before introducing the robust amplitude estimation algorithm we briefly review the standard sampling estimation method.
In the simplest setting, standard sampling is used in VQE to estimate expectation values of Pauli strings.
For a Hamiltonian decomposed into a linear combination of Pauli strings $H=\sum_j \mu_j P_j$ and ``ansatz state'' $\ket{A}$, the energy expectation value is estimated as a linear combination of Pauli expectation value estimates
\begin{align}
\label{eq:energyestimator}
    \hat{E}=\sum_j \mu_j \hat{\Pi}_j,\\
    \textup{Var}(\hat{E})\leq \varepsilon^2_{\textup{chem. acc.}}
\end{align}
where $\hat{\Pi}_j$ is the estimate of $\langle A|P_j|A\rangle$.
For a given Pauli operator $P$, the standard sampling estimation procedure is as follows: prepare $\ket{A}$ and measure operator $P$ receiving outcome $d=0,1$; repeat $M$ times, receiving $k$ outcomes labeled $0$ and $M-k$ outcomes labeled $1$; estimate $\Pi=\langle A|P|A\rangle$ as $\hat{\Pi}=\frac{k-(M-k)}{M}$.
In the case that one has some prior information about the value of $\Pi$, we can use a Bayesian inference variant of the above estimation process. In this case, expectation values are modeled as binomial distributions with beta priors, such that the measurement process can be assimilated as updating the distribution based on new measurements according to Bayes rule.
This approach is referred to as \emph{Bayesian VQE} (BVQE), and is described in \cite{McClean2016}.

As determined in \cite{gonthier2020identifying}, reaching a high-accuracy energy estimate with standard sampling requires too many independent measurements for VQE to compete with state-of-the-art classical quantum chemistry methods.
This is due to the large constant of proportionality $K$ relating the estimation runtime T to the target accuracy $\varepsilon$,
\begin{align}
    T= MC = \frac{CK}{\varepsilon^2},
\end{align}
where $M$ is the number of measurements and $C$ is the time cost of state preparation and measurement.
This large proportionality constant of $CK$ is the source of the measurement problem.
As described in \cite{gonthier2020identifying}, state-of-the-art methods reduce the value of $K$ but still fall several orders of magnitude short.
We refer to this issue as \emph{the measurement problem}.
This finding illuminates an obstacle for using VQE as a practical problem solving method and motivates the need for methods which reduce the runtime of estimation more dramatically.

\subsection{Robust amplitude estimation}
\label{subsec:rae}
The robust amplitude estimation algorithm serves to speed up the estimation of expectation values. A detailed description of this method can be found in the following reference \cite{wang2021minimizing}.
We introduce the RAE algorithm as a solution to the measurement problem discussed in the previous section.
In contrast to the estimation method typically used in VQE, RAE enables a reduction of estimation runtime proportional to improvements in the quality of the quantum hardware.
Accordingly, we expect that for quantum devices of sufficient quality, we can use the RAE algorithm to carry out energy estimation in a reasonable amount of time.

The robust amplitude estimation algorithm is used to speed up the estimation of each expectation value
\begin{align}
\Pi = \cos{\theta} =\bra{A} P \ket{A},
\label{eq:defPi}
\end{align}
where $\ket{A}=A\ket{0^n}$ in which $A$ is the ansatz circuit, $P$ is an $n$-qubit Hermitian operator with eigenvalues $\pm 1$, and $\theta=\arccos{\Pi}$ is introduced to facilitate Bayesian inference. 
The only substantially new circuit operation that is required for this method is a reflection about the initial state $R_0=\mathbb{I} -2\ket{0}\! \bra{0}^{\otimes N}$.

\begin{figure}[!ht]
\center
\includegraphics[width=.95\columnwidth]{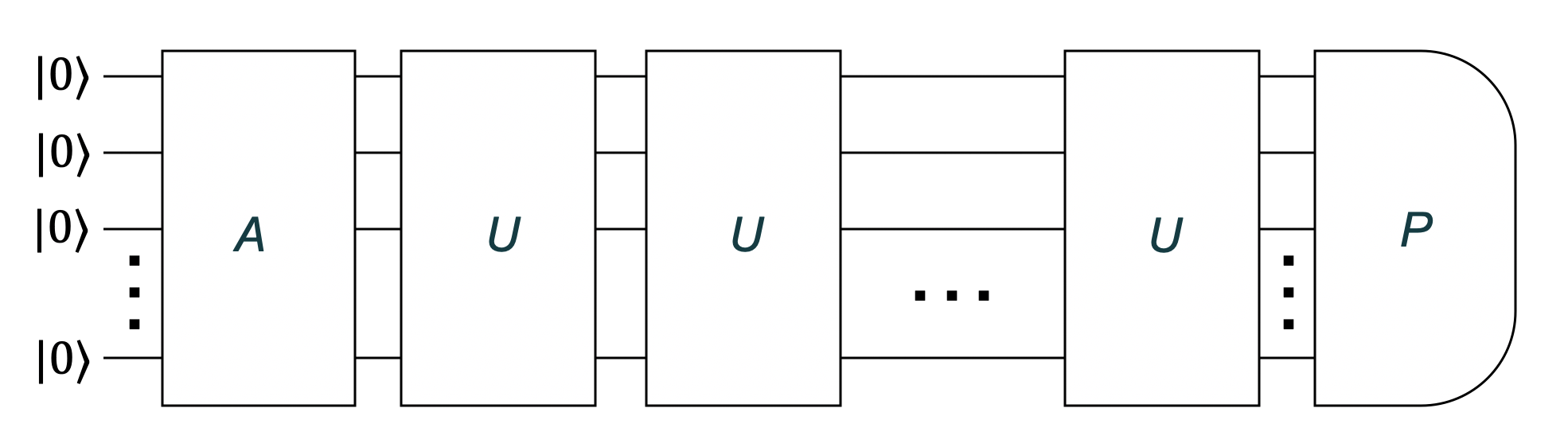}
\caption{This figure depicts the operations used for generating measurement outcomes in the robust amplitude estimation algorithm.
$A$ is the state preparation circuit, $P$ is the observable of interest (the final circuit operation indicating it's measurement), and $U$ is the Grover iterate comprised of $AR_0A^{\dagger}P$, with $R_0$ the reflection about the state $\ket{0}^{\otimes N}$.} 
\label{fig:circuit_diagram}
\end{figure}
RAE uses the quantum circuit shown in Figure~\ref{fig:circuit_diagram} to generate measurement outcomes as follows:
prepare the ansatz state $\ket{A}=A\ket{0^n}$, apply 
$L$ RAE circuit layers $U=AR_0A^{\dagger}P$, and then measure the Pauli observable $P$.
In the noiseless setting, the likelihood of the outcomes $d=0,1$ depends on the parameter of interest $\Pi=\cos\theta$ as
\begin{align}
\mathbb{P}(d|\theta; L)=\dfrac {1}{2}\left( 
1+(-1)^d \cos((2L+1)\theta) \right).
\end{align}
RAE uses the outcomes from a sequence of such measurements to infer the true value of $\theta$, and hence $\Pi=\langle A|P|A\rangle$.
This inference can be implemented in a variety of ways including filtering techniques \cite{wiebe2016efficient}, numerical maximum likelihood estimation \cite{tanaka2020amplitude}, and adaptive grid refinement \cite{tipireddy2020bayesian}.
At the core of each of these methods is the Bayes update rule, whereby a prior distribution $p(\theta)$ capturing initial beliefs about the parameter of interest is updated to a posterior distribution $p(\theta|d)$ by multiplying by the likelihood function and dividing by the model evidence,
\begin{align}
p(\theta|d)=\frac{\mathbb{P}(d|\theta)p(\theta)}{\int d\theta \mathbb{P}(d|\theta)p(\theta)}.
\end{align}
In essence, RAE reduces the estimation runtime by drawing measurement outcomes whose likelihoods depend sensitively on the parameter of interest.

In practice, quantum computation is subject to errors.
These errors derive from several sources, including decoherence to the thermal environment and limitations on the calibration of quantum gates.
Such errors effect the relationship between the parameter of interest $\Pi$ and the likelihoods of observed data; the actual likelihood function differs from the idealized likelihood function.
To accommodate this, the robust amplitude estimation algorithm incorporates a model for the impact of noise on the inference process.
We employ the noise model of \cite{wang2021minimizing} to account for the effect of error in RAE.
The purpose of the noise model is to predict how errors in the circuit implementation influence the likelihood function from which measurement outcomes are drawn.
By composing $L$ noisy RAE circuit layers and measuring the Pauli observable, we model the resulting likelihood function using the \emph{exponential decay} model,
\begin{align}
\label{eq:likelihoodfunction}
    \mathbb{P}(d|\theta, \lambda, \overline{p}; L)=\dfrac {1}{2}\left( 
1+(-1)^d \overline{p}\exp{(-\lambda L)}\cos((2L+1)\theta) \right),
\end{align}
where $\overline{p}$ denotes the initial ansatz state preparation and measurement error and $\lambda$ denotes the exponential decay parameter.
These two parameters can either be fixed (possibly learned from prior experiments) or can be treated as \emph{nuisance} parameters to be learned during the inference process \cite{royall2000probability}. 

In \cite{wang2021minimizing}, the authors propose a model for the runtime (converted here to be measured in total number of queried RAE circuit layers) 
for robust amplitude estimation to reach a target mean squared error $\varepsilon^2$, given the noise parameters $\overline{p}$ and $\lambda$:
\begin{align}
\label{eq:runtimemodel}
t_{\varepsilon} \approx \frac{e^2}{e-1}\frac{e^{-\lambda}}{2\overline{p}^2}\left(\frac{ \lambda }{\varepsilon^2} + \frac{1}{\sqrt{2}\varepsilon} +\sqrt{\left(\frac{\lambda}{\varepsilon^2}\right)^2+\left(\frac{2\sqrt{2}}{\varepsilon}\right)^2}\right),
\end{align}

The model shows an interpolation between the traditional scalings known as the shot-noise-limit scaling $O(1/\varepsilon^2)$ and the Heisenberg-limit scaling $O(1/\varepsilon)$. 
As described in the following section, one of the contributions of this work is the analysis of simulation data which validates this model with respect to a weaker set of assumptions than those used to derive the model.
As explained in the following section, this weaker set of assumptions still assumes that the influence of error is perfectly modeled by the exponential decay model likelihood function, leading to over-optimistic conclusions of algorithm performance.
Importantly, this suffices for our purposes because we aim to understand the minimal resources necessary, but not sufficient, for achieving quantum advantage.


\section{Methods}
\label{sec:methods}
The main objective of this work is to predict the resources that are necessary, but not necessarily sufficient, for achieving quantum advantage for the problem of estimating molecular ground state energies.
A critical bottleneck in using quantum variational methods to determine the ground state energy is the large number of statistical samples needed for accurate estimates\cite{gonthier2020identifying}.
Thus, a necessary condition for achieving quantum advantage for the ground state energy estimation problem is to carry out the estimation task in a reasonable amount of time. The methods we detail in this section aim to predict the total duration of time needed to ensure an energy estimation to within chemical accuracy for the molecules of interest.

\subsection{Runtime prediction strategy}
\label{subsec:methods-summary}
The task which we analyze is the estimation of the energy expectation value $\bra{A}H\ket{A}$ using RAE, where $\ket{A}=A\ket{0}^{\otimes N}$ is the ansatz state.
The analysis of this task using VQE was described in \cite{gonthier2020identifying}.
Each molecular Hamiltonian of interest is converted into a weighted sum of Pauli terms as $H=\sum_j \mu_j P_j$.
The estimation method first estimates the expectation value of each Pauli term individually as $\hat{\Pi}_j$ and then takes the weighted sum of such estimates as the final energy expectation value estimate according to Equation (\ref{eq:energyestimator}). 
We consider the case of having a target accuracy $\varepsilon$ for the estimate of the final energy, and we wish to minimize the total time required to achieve this target accuracy.
Excluding the possibility of estimating Pauli expectation values in parallel across different QPUs from our analysis, we must choose how to optimally allocate time to each estimation so as to minimize the total time. In particular, we should spend proportionally more time, and thus achieve a better accuracy $\varepsilon_j$, on terms with larger coefficients $\mu_j$.

For a single-term estimate $\hat{\Pi}_j$, the total runtime depends on the number of ansatz and phase flip reflection operations used for each estimate and the time taken to implement each of these operations.
Since each RAE circuit consists of repeated layers of $U = A R_0 A^\dagger P_j$, we will first count the total number of such layers used in the estimation process, and then determine the time needed to implement each layer.
That time depends on the compilation of the component operations into elementary logical gates. Finally, the time needed to implement each logical gate depends on the quantum error correction resources used and on the runtime of each underlying physical gate.
Qualitatively, a larger target fidelity of the gate requires a larger time overhead for its fault-tolerant implementation.
Finally, the modality used for the quantum computer can have a significant impact on the time needed for each elementary gate; as a rule of thumb, elementary gates implemented with superconducting-qubit quantum computers tend to be several orders of magnitude faster than those of ion trap quantum computers.
Putting all of this together, we arrive at an estimate of the time required to reach a target accuracy in the energy estimate.

We introduce some notation and outline this quantitative strategy. 
The inputs to the runtime prediction are
\begin{itemize}
    \item $H$: Hamiltonian
    \item $\varepsilon$: Target accuracy, where $\varepsilon^2=\mathbb{E}(\hat{E}-\langle H\rangle)^2$
    \item $N$: Number of logical qubits
    \item $r_g$: Elementary physical gate error rate
    \item $\overline{r}_g$: Elementary logical gate error rate
    \item $T_g$: Elementary physical gate time
\end{itemize}
These inputs determine a number of dependent quantities, which are used to carry out the final runtime prediction.
The steps of this process are enumerated as follows:
\begin{enumerate}
    \item Validate the runtime model of Equation (\ref{eq:runtimemodel}) for single-term estimation to target accuracy $\varepsilon_j$: $T_j = \tau_l \cdot t_j(\varepsilon_j, \lambda, \overline{p})$, where $\tau_l$ is the duration of a RAE circuit layer and $t_j$ is the total count of the number of RAE layers queried (see Section \ref{subsec:rae} for $\varepsilon_j$, $\lambda$, and $\overline{p}$).
    \item Establish a method for allocating runtime among terms, amounting to determining optimal target accuracies for each term $\varepsilon_j^*$.    
    \item Determine circuit depths of the ansatz and phase flip operations $D_A$ and $D_R$, respectively. 
    Then determine the required logical gate error rate $\overline{r}_g$ from $e^{-\lambda}=(1-\overline{r}_g)^{(2 D_A + D_R)N/2}$. 
    The layer runtime is determined from $\tau_l=(2D_A+D_R)\tilde{T}_g$.
\item Determine fault-tolerant overhead $F(\overline{r}_g, r_g)$ needed to achieve logical gate error rate $\overline{r}_g$ with physical gate error rate of $r_g$, giving logical gate time $\tilde{T}_g=F(\overline{r}_g, r_g)T_g$.    
\end{enumerate}
With these in place the final expression for the runtime to target accuracy is 
\begin{align}
\label{eq:ft_overhead}
    T = \tau_l(D_A, D_R, \tilde{T}_g)\sum_jt_j(\varepsilon_j^*, \lambda, \overline{p}).
\end{align}
We note that the time needed for carrying out the estimates in parallel is simply the maximum of the times among the individual terms,
\begin{align}
    T_{\|} = \tau_l(D_A, D_R, \tilde{T}_g)\max_jt_j(\varepsilon_j^*, \lambda, \overline{p}).
\end{align}
The methods used for the step of validating the runtime model are reported in Appendix \ref{sec:single-term-runtime}, and the corresponding results in Appendix \ref{subsec:validation_results}. Appendix \ref{sec:compilation} describes the compilation model and the assessment of circuit characteristics; Appendix \ref{sec:shot_allocation} describes the method for allocating shots across the Pauli expectation values; and Appendix \ref{sec:fault_tolerance} describes the fault-tolerance cost model used in the final results of Section \ref{sec:results}.
In the remainder of this section we detail the methods used to make the runtime predictions for standard sampling and RAE.

\subsection{Resource estimation methods}
\label{subsec:meth_runtime_prediction}

In this subsection we describe the methods for generating energy estimation runtime predictions for standard sampling and RAE.

The problem instances defined in \cite{gonthier2020identifying} comprise a set of small hydrocarbons for which combustion energies should be calculated to an accuracy comparable to that achievable in experiments. For this purpose, between 104 and 260 qubits are necessary due to the large basis sets involved. Since obtaining and manipulating Hamiltonians for problems of this size is cumbersome with the currently available software, we instead generated two series of Hamiltonians for up to 80 qubits for each molecule, and used the results to extrapolate to the large qubit numbers. The two series of Hamiltonians are generated with two different orbital types, i.e. two different discretizations for the problem, like in \cite{gonthier2020identifying}. As detailed in \cite{gonthier2020identifying}, we built Hamiltonians so that the number of qubits used is an integer multiple of the number of active electrons, to facilitate extrapolation. For example, for H$_2$O with 8 active electrons, we built Hamiltonians with 16, 24, 32, 40, 48, 56, 64, 72 and 80 qubits. The Hamiltonian with only 8 qubits is omitted since it would trivially yield the mean-field Hartree-Fock energy. 

To obtain the predicted runtime for RAE, we use the following steps:
\begin{enumerate}
    \item Estimate the total number of Ansatz queries required by RAE to reach chemical accuracy for each Hamiltonian, based on the runtime model validated in Appendix \ref{sec:single-term-runtime} and on the allocation detailed in Appendix \ref{sec:shot_allocation}. This was done for final RMSEs of 10$^{-3}$ and 10$^{-4}$ Ha and for circuit layer fidelities $e^{-\lambda}$ comprised between $(1 -10^{-3})$ and $(1 - 10^{-6})$ on an approximately logarithmic scale.
    \item For a fixed molecule, orbital type, RMSE and layer fidelity, the number of Ansatz queries was fitted using SciPy as a function of the number of qubits with $aN^b + c$ where $a$, $b$, and $c$ are fitting coefficients. In the few cases where only two data points were available, the coefficient $c$ was fixed to zero.
    \item Plug in the appropriate number of qubits (between 104 and 260) in the function resulting from the fit to estimate the number of Ansatz queries necessary to reach chemical accuracy.
    \item Convert the number of Ansatz queries to runtimes in seconds using Equation (\ref{eq:ft_overhead}).
\end{enumerate}
 
The runtime predictions for standard sampling follow those of \cite{gonthier2020identifying} with a few modifications.
Compared to \cite{gonthier2020identifying}, we consider the runtime of standard sampling energy estimation for varied gate error rates.
Accordingly, we introduce two competing factors in the runtime predictions.
The gate error rates are decreased using quantum error correction; as mentioned above, improving gate error incurs a time overhead that we factor into the runtime predictions.
This reduction in gate error rate is helpful, though, in that it reduces the degree to which any error mitigation technique will adjust the expectation value estimates.
We will assume an idealistic error mitigation technique that simply rescales the expectation value estimates so as to invert the attenuation factor in the observed expectation value $\langle P_i\rangle_{obs} = f\langle P_i\rangle_{ideal}$, where $f$ is the circuit fidelity.
An example of such an estimator is probabilistic error cancellation \cite{mari2021extending}.
In rescaling the observed expectation value by $1/f$, the variance in the estimate is scaled by $1/f^2$ (c.f. Eq. 15 in \cite{mari2021extending}). 
Accordingly, we model the error mitigation overhead as a factor in the standard sampling energy estimation runtime 
\begin{align}
    \frac{1}{f^2}=\frac{1}{(1-\overline{r}_G)^{D_AN}},
\end{align}
where we take the circuit fidelity to be the product of the logical gate fidelities $1-\overline{r}_G$ of all $D_AN/2$ gates in the ansatz circuit. 
Note that we have made the optimistic assumption that the readout error is negligible compared to that of the gate error.
Hence, to obtain the predicted standard estimation runtimes at large qubit numbers, we follow these steps:
\begin{enumerate}
\item Compute $K$ for each Hamiltonian according to the method presented in \cite{gonthier2020identifying}, except that \emph{no grouping} of Pauli terms or variance reduction technique is applied to the Hamiltonians.
\item For a fixed molecule and orbital type, fit $K$ to $aN^b+c$ where $a$, $b$, $c$ are fitting coefficients and $N$ the number of qubits. $c$ is set to zero if only two data points are present. 
\item Evaluate $M=K/\varepsilon^2$ with $K$ obtained from the formula fitted above applied to large qubit numbers, for $\varepsilon$ of 10$^{-3}$ and 10$^{-4}$ Ha.
\item Compute the error mitigation overhead. The logical gate error rate is chosen to be consistent with the RAE circuit fidelities chosen above. Factor the error mitigation overhead with the execution time obtained from $D_{A}$ and the logical gate execution time $\tilde{T}_g$ that includes the appropriate error correction overhead.
\end{enumerate}

In Section 
\ref{sec:results} 
we present plots and summarize the findings of these extrapolations.

\section{Results}
\label{sec:results}

\begin{figure*}[t!]
     \centering
     \begin{subfigure}{}
         \centering
         \includegraphics[width=0.48\textwidth]{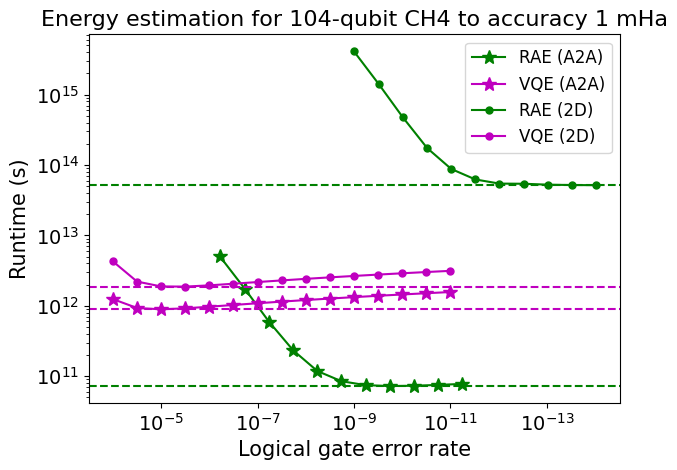}
         \label{fig:CH4}
     \end{subfigure}
     \hfill
     \begin{subfigure}{}
         \includegraphics[width=0.48\textwidth]{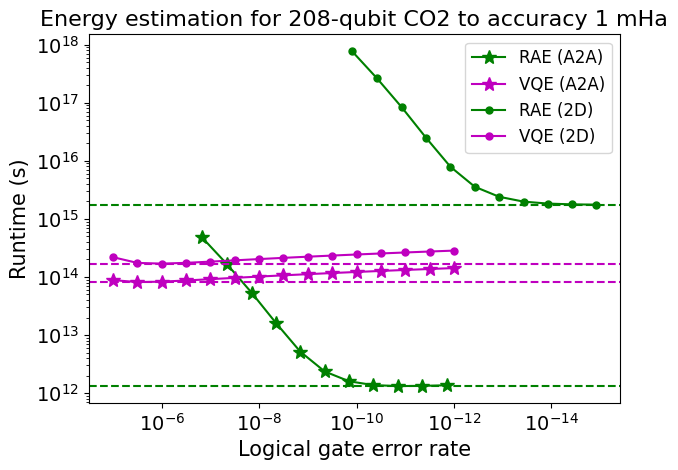}
         \label{fig:CO2}
     \end{subfigure}
        \caption{These figures compare the predicted runtimes of standard sampling (magenta), as typically used in VQE, and robust amplitude estimation (green) for estimating the energies of the CH$_4$ and CO$_2$ molecules to within accuracy of 0.001 Hartrees as the logical gate error rate is improved through error correction.
        The star and dot symbols indicate the use of compilation from all-to-all (A2A) connectivity and two-dimensional (2D) connectivity, respectively.
        The horizontal dotted lines mark the minimal predicted runtimes for each method and connectivity.
We have chosen the CH$_4$ and CO$_2$ molecules because they yield the smallest and largest speedups of RAE over standard sampling, respectively.
For both molecules, using RAE with A2A connectivity yields the lowest predicted runtime. However, the predicted runtimes are still too high to be practical.}
\label{fig:runtime_plots}
\end{figure*}

We now describe the results of the runtime prediction described in Section \ref{subsec:meth_runtime_prediction}.
These results are depicted in detail for two specific molecules in Figure \ref{fig:runtime_plots} and summarized for the full set of molecules in Table \ref{tab:runtime_table}.

Figure \ref{fig:runtime_plots} shows the predicted runtimes for the two estimation methods (standard sampling of VQE and robust amplitude estimation) as a function of logical gate error rate.
We have chosen the molecules CH$_4$ and CO$_2$ to represent the smallest ($13\times$) and largest ($64\times$), respectively, relative improvements of RAE over standard sampling.
The figure shows the result of how reducing the logical gate error rates affords the RAE algorithm to run deeper quantum circuits, increasing the degree of quantum amplification and subsequently reducing the runtime of the estimation task.
The shape of the runtime vs 
error rate curve reveals an important phenomenon: for a target estimation accuracy, the runtime has a minimum at an optimal logical gate error rate.
We observe the presence of a minimum for both RAE and standard sampling.
In the case of RAE, this phenomenon is due to the balance of increased quantum amplification with the cost of error correction overhead.
In the case of VQE, the phenomenon is due to the balance of reduced cost of error mitigation overhead with the cost of error correction overhead.

This figure also compares the predicted runtimes of estimation when using compilation to different device connectivities: all-to-all (A2A) and two-dimensional (2D).
We observe that the best predicted runtimes are achieved by RAE when all-to-all compiling is used. 
When the operations are compiled according to a 2D connectivity, the predicted runtime of RAE is larger than that of VQE with any connectivity; this is due to the phase flip operation requiring many gates when compiled according to a 2D connectivity (see Table \ref{tab:compilation}).

\begin{table*}
\begin{tabular}{lrrrrrrrrrrr}\\
\hline Molecule &     C$_2$H$_2$ &     C$_2$H$_4$ &     C$_2$H$_6$ &    C$_2$H$_6$O &     C$_3$H$_4$ &     C$_3$H$_6$ &     C$_3$H$_8$ &      CH$_4$ &     CH$_4$O &      CO$_2$ &      H$_2$O \\ 
\hline\hline 
Number of logical qubits      & 130 & 156 & 182 & 260 & 208 & 234 & 260 & 104 & 182 & 208 & 104 \\ 
Number of VQE physical qubits & 43,900 & 61,200 & 71,300 & 117,000 & 81,500 & 105,000 & 117,000 & 35,200 & 71,300 & 81,500 & 35,200 \\ 
Number of RAE physical qubits & 162,000 & 195,000 & 228,000 & 325,000 & 260,000 & 292,000 & 352,000 & 120,000 & 228,000 & 281,000 & 130,000 \\
\hline
Optimal VQE gate error rate   & 1e-5 & 4e-6 & 4e-6 & 2e-6 & 4e-6 & 2e-6 & 2e-6 & 1e-5 & 4e-6 & 4e-6 & 1e-5 \\
Crossover RAE gate error rate & 4e-8 & 3e-8 & 2e-8 & 9e-9 & 1e-8 & 1e-8 & 9e-9 & 6e-8 & 2e-8 & 1e-8 & 6e-8 \\ 
Optimal RAE gate error rate   & 4e-11 & 3e-11 & 6e-11 & 3e-11 & 4e-11 & 4e-11 & 9e-12 & 2e-10 & 6e-11 & 1e-11 & 6e-11 \\
\hline
Estimation runtime VQE (10$^{9}$s)    & 1,700 & 3,900 & 22,100 & 477,000 & 98,200 & 125,000 & 189,000 & 910 & 32,700 & 83,900 & 1,400 \\ 
Estimation runtime RAE (10$^{9}$s)    & 70 & 130 & 1,100 & 20,000 & 5,100 & 6,200 & 7,000 & 72 & 1,800 & 1,300 & 41 \\ 
Runtime ratio (VQE/RAE)       & 25 & 30 & 20 & 25 & 19 & 20 & 27 & 13 & 18 & 64 & 34 \\ 
\hline
\end{tabular} 
\centering
\caption{This table shows the predictions of resources needed to estimate the energy of the prepared ground state using standard sampling (denoted as VQE) or RAE to below chemical accuracy (1.0mHa<1.3mHa) for each molecule (represented using canonical orbitals).
For both standard sampling and RAE (for each molecule), we choose the logical gate error rate so that the runtime is minimized, and we report it as the optimal error rate. In addition, for RAE we indicate the crossover error rate, at which RAE yields lower runtimes than standard sampling. 
The phase flip operation used in RAE is compiled assuming an all-to-all connectivity and we assume surface code cycle times of 1$\mu$s.
}
\label{tab:runtime_table}
\end{table*}

Table \ref{tab:runtime_table} shows physical and logical qubit resources, gate error rates, and runtime data for each of the eleven molecules studied.
We compare these resources and runtimes for standard sampling and RAE. In each case, the data is presented for the optimal logical error rate (minimal runtime) unless otherwise specified.
In the case of standard sampling, the number of physical qubits (accounting for error correction overhead) ranges from tens of thousands to roughly a hundred thousand qubits.
RAE can take advantage of further reducing the logical error rate. This comes at the cost of an additional qubit overhead that is three to four times that of standard sampling.
This additional overhead results in more efficient error correction, bringing the logical error gate for optimal RAE operation five to six orders of magnitude below that of standard sampling.
This reduction in logical gate error rate enables a factor of 13 to 64 reduction in the RAE runtime compared to standard sampling.

Despite the predicted runtime improvements of RAE, we observe that these times are still too high to be practical; the lowest predicted runtime is more than \emph{one millenium}. However, like standard sampling, this estimation method is highly parallelizable; running on multiple quantum processing units gives a proportional reduction in runtime.
Furthermore, we highlight once again that these runtime estimates do not include any grouping or variance reduction methods. In the case of standard sampling, such methods can reduce the estimated runtime by three to five orders of magnitude. 
However, RAE operates quite differently from standard sampling, and thus it remains to be seen if a similar improvement can be obtained and how methods used in standard sampling can be adapted. 

\section{Discussion and Outlook}

\label{sec:outlook}
This work contributes to identifying the most promising near-term methods for achieving quantum advantage in molecular simulation.
In previous work \cite{gonthier2020identifying}, we identified the ``measurement problem'' as a bottleneck in the variational quantum eigensolver (VQE) algorithm, making the subroutine of energy estimation prohibitively slow even for small molecules of industrial relevance.
Here, we have investigated the extent to which quantum amplification helps to reduce the runtime needed to estimate the energy expectation value in VQE.
Specifically, we carried out runtime predictions for the energy estimation subroutine when using robust amplitude estimation (RAE) \cite{wang2021minimizing}.
We used a performance model for RAE (see Appendix \ref{subsec:validation_results}) to carry out runtime predictions for a range of system sizes and molecules.
We then extrapolated these results to the larger system sizes of interest.

Using a coarse approach to fault-tolerant compilation and making optimistic assumptions about the operation speeds of the fault-tolerant architecture, we arrive at runtime estimations in terms of number of seconds.
We find that using RAE gives a 13 to 64 factor speedup over standard sampling, with RAE requiring just a few times as many physical qubits as VQE from the error correction overhead.
Although the predicted runtimes for both methods are still too high to be of practical value, we note that they were obtained without any grouping of the Hamiltonian terms or variance reduction techniques. In the case of standard sampling, such methods can reduce the predicted runtime by up to 5 orders of magnitude. 
Some grouping methods are compatible with RAE, but it remains to be seen whether the associated cost will yield practical methods. We leave this investigation to future work. 

Beyond Hamiltonian grouping techniques, we discuss several other ways in which predictions with more reasonable runtimes might be obtained:
\begin{enumerate}
\item Circuit depth reduction: 
Any reduction in circuit depth saves on time and error; both of which reduce the runtime of RAE.
It is possible to reduce circuit depth through improved compilation schemes (e.g. in the phase flip \cite{baker2019decomposing}) or improved ansatz circuits \cite{kottmann2021optimized}.
Furthermore, the cost of the input-state reflection in RAE can be greatly reduced by exploiting symmetry in the ansatz circuit and Pauli operators. In particular, for number-conserving ansatz and diagonal Pauli operators, it is possible to reduce the costly reflection to a $k$-qubit operation (where $k$ is the number of electrons). We estimate the time savings here to be roughly an order of magnitude.
\item Improved fault-tolerant resource estimation: the current resource accounting is quite coarse. A fine-grained resource estimation could lead to either more or less optimistic predictions. However, we did not account for optimizations in the fault-tolerant compilation. For example, one can optimally allocate spacetime volume between qubit counts and time (see, e.g. Appendix C of \cite{kim2021fault}).
Moreover, recent advances in quantum error correction \cite{panteleev2021asymptotically, tremblay2021constant, cohen2021low} may eventually lead to further reductions in predicted runtimes.
\item Quantum-apt application instances: 
alternative sets of molecules that are still of industrial relevance may serve as better candidates for achieving quantum advantage.
Preliminary resource estimations indicate that, for the set of molecules considered, even modern quantum phase estimation techniques take on the order of hours to days to run. 
We believe that a critical effort towards realizing quantum advantage will be the development and identification of problem instances which are ``easy'' for a quantum computer and ``hard'' for state-of-the-art classical methods, while still being of practical value.
\end{enumerate}


Thus far, we have considered robust amplitude estimation as a tool to speed up the energy estimation subroutine of VQE.
Yet there are other aspects of VQE which stand to be improved.
In the outer loop of the VQE algorithm, a classical optimization process is used to find the ansatz parameters for which the corresponding circuit well approximates ground state preparation.
This process can be improved by designing better ansatz circuits \cite{Dallaire2019, kottmann2021optimized} and finding more effective methods for parameter optimization \cite{sim2021adaptive}.
Recent work introduced the concept of state preparation boosters \cite{wang2022state} as a method to reliably increase ground state overlap at the cost of using deeper quantum circuits.

Given the remaining challenges for VQE, it is likely that additional quantum algorithm methods will be needed to solve problems of industrial value \cite{lin2022heisenberg, zhang2021computing, wan2021randomized}.
This is consistent with the perspective that VQE might be used to get a ``head start'' in the state preparation subroutine for more powerful quantum algorithms, i.e. the output state of the VQE calculation provides a rough approximation of the ground state that can be used as an input for another quantum algorithm.

The next decade is sure to bring the value of quantum computing more clearly into view.
We hope this work will help guide the community towards identifying the first quantum computing use cases in quantum chemistry and we expect insights from this benchmark study to inform future benchmarks for molecules that show promise for early quantum advantage.

\section*{Acknowledgements \label{Acknowledgement}}
The authors would like to acknowledge support from BP. The authors also acknowledge insightful scientific discussions and suggestions from the team of scientists and engineers at Zapata Computing. Numerical results were generated using the Orquestra\textregistered{} platform by Zapata Computing Inc.

\appendix

\section{Validation of single-term estimation runtime model}
\label{sec:single-term-runtime}

In this section of the appendix we present the methods and results for validating the single-term estimation runtime model of Equation (\ref{eq:runtimemodel}).

\subsection{Methods}
\label{subsec:single-term-runtime-methods}
An accurate theoretical analysis of RAE's performance is difficult to obtain due to the adaptivity of the algorithm. In each step of inference, $L$ is chosen to maximize the expected gain in Fisher information per time spent for that sample.
However, by making several approximations and assumptions, \cite{wang2021minimizing} arrived at upper and lower bounds on the runtime to target estimation.
While these bounds were derived for the case of using ``engineered likelihood functions'' (c.f. \cite{wang2021minimizing}), we will find that the bounds mostly capture the performance of estimation with Chebyshev likelihood functions, which we are simulating.
Our goal is to test the accuracy of these bounds through simulation of the inference process, while relaxing some assumptions used in the runtime model.
The question we aim to answer is: how accurate are the runtime bounds in the case where the true sample rates $\mathbb{P}(d|\theta; L)$ \emph{match} those of the likelihood function $\mathbb{P}(d|\theta,\lambda; L)$ (we assume $\overline{p} = 1$)?
In practice, we expect a discrepancy between the true sample rates and the likelihood function used for inference because of the discrepancy between our noise model and actual noise.
This discrepancy will, in general, make the expectation value estimates less accurate, leading to worse performance of the algorithm.
Accordingly, we will view our results as roughly establishing a lower bound on performance: we expect estimation runtimes in practice to be longer than the ones we obtain.

The three main assumptions used in the simulations are:
\begin{enumerate}
    \item The effect of noise is described by the exponential decay model of Equation (\ref{eq:likelihoodfunction}) with a known decay parameter $\lambda$.
    \item The decay parameter $\lambda$ is determined by the number of effective two-qubit gates in a single Grover iterate (accordingly, $\lambda$ is independent of the Pauli term involved in the Grover iterate).
    \item The duration of the estimation process is determined by the cumulative duration of the circuit implementation time.
\end{enumerate}
Regarding assumptions 1 and 2, in practice the likelihood function is simply a model for the relationship between the parameter of interest and the outcome likelihoods.
Therefore, our runtime model does not fully capture the case of likelihood function inaccuracies.
However, the present evaluation provides a baseline for further analysis including such inaccuracies.

Regarding assumption 3, the total estimation time in practice will include measurement time and a latency between each measurement due to subsequent re-initialization.
However, as the duration of the quantum circuits increases, the relative proportion of time spent on measurement and latency will decrease.
Because we will be considering circuits of considerable depth in the regime where some quantum error correction is used, we will take this measurement and latency time to be negligible.

In our numerical experiments, we run simulated inference under a number of different settings and estimate the median error over many trials.
In each trial, the prior distribution is chosen as a Gaussian distribution over the phase angle $\theta = \arccos(\Pi)$ in a randomized fashion as follows.
The standard deviation of the prior is set to $\sigma=0.01$ and the mean is drawn from a Gaussian distribution centered around the true phase angle with standard deviation $\sigma=0.01$.
This choice reflects a practical strategy for energy expectation value estimation: means of the prior would be set to values derived from the best classical methods (in this case, coupled cluster) and the prior standard deviations initialized to be larger than the typical errors for the classical method used.
We have found that 0.01 is typically larger than the error between the true ground state expectation values and the classical method expectation value.

We simulate the adaptive inference process for robust amplitude estimation using Chebyshev likelihood functions as described in \cite{wang2021minimizing}.
We vary the true expectation value and the layer fidelity as:
\begin{itemize}
    \item $\Pi$ = [0, 0.15, 0.3, 0.45, 0.6, 0.75, 0.9]
    \item $e^{-\lambda}$ = [0.9, 0.99, 0.999, 0.9999, 0.99999, 0.999999]
\end{itemize}
Note that $\pm\Pi$ should yield the same performance due to symmetry of the likelihood function; accordingly, we only consider $\Pi\ge0$.
Ultimately we are aiming to understand the relationship between accuracy and runtime.
For the higher-fidelity trials, the change in accuracy per additional time is far greater than the lower-fidelity trials.
Accordingly, we use fewer steps of Bayesian inference in the higher-fidelity trials.
For each setting, we simulate between $274-354$ trials and track the error between the current estimate and the true value at each step of Bayesian inference.

The quantity we use to assess the performance of the estimation process is the mean squared error.
However, direct use of the mean squared error does not reflect the quality of the estimators.
This is due to occasional estimates far from the true value.
Accordingly, we have chosen to plot an estimate of the mean-squared error of the 10th percentile of the estimates.
That is, we first exclude the worst $10\%$ of the estimates and then compute the sample MSE from the remaining estimates.
In the following section, we compare these adjusted MSEs to the MSEs predicted by the runtime model.

\subsection{Results}
\label{subsec:validation_results}
\begin{figure*}
\center
\includegraphics[width=18cm]{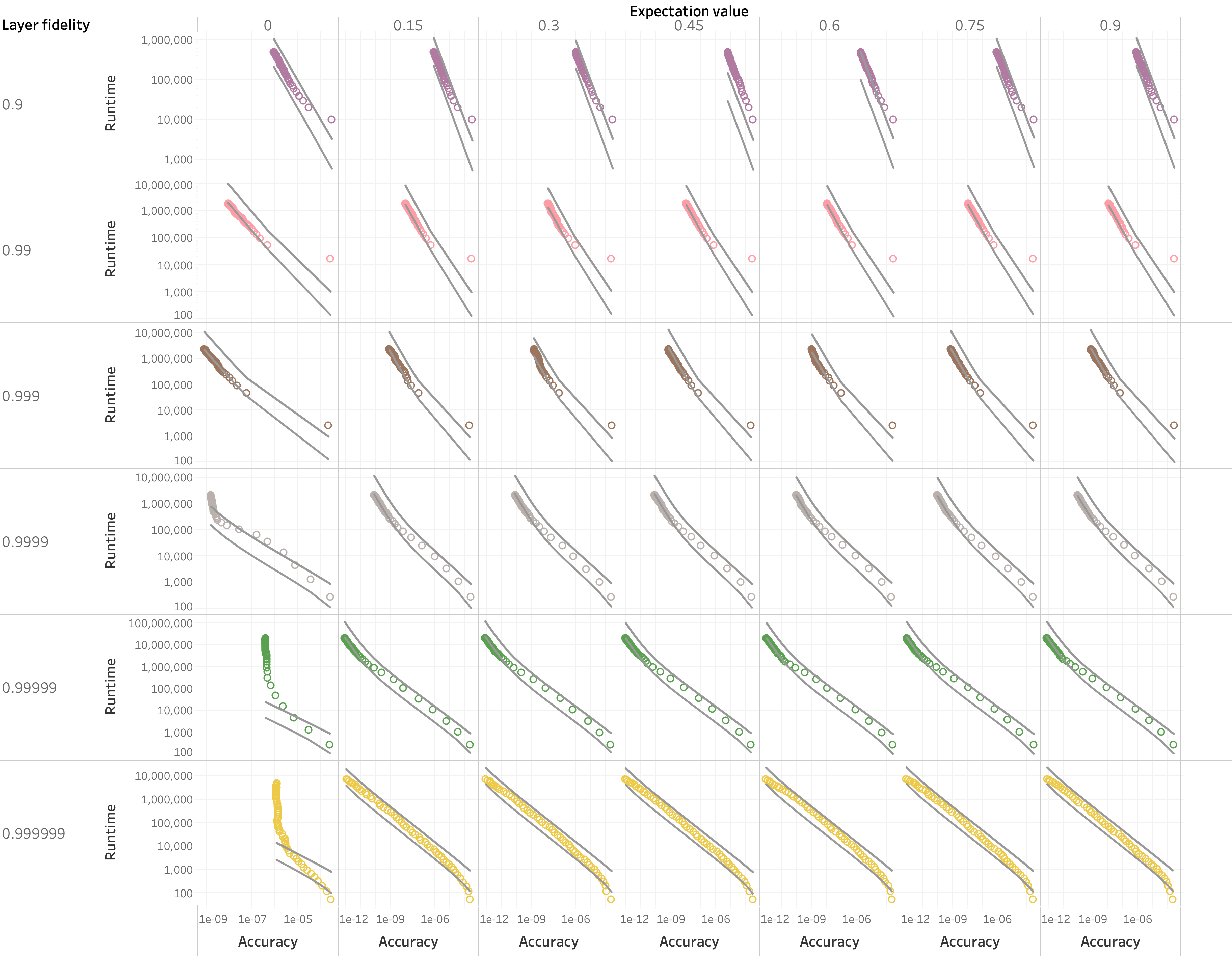}
\caption{This figure compares the runtime model of Eq. (\ref{eq:runtimemodel}) to the estimation runtimes found in simulation. Each inset plot shows the accumulated runtime in terms of total number of circuit layers queried (computed as the average runtime of the best 90$\%$ of trials) as a function of the empirically estimated estimation accuracy (computed as the average squared error of the best 90$\%$ of trials).
Each row corresponds to a different circuit layer fidelity and each column corresponds to a different true expectation value.
For each inset, the upper and lower bounds of the runtime model are plotted as grey curves.
} 
\label{fig:model_validation}
\end{figure*}

We present the results validating the runtime model of Equation (\ref{eq:runtimemodel}) in Figure \ref{fig:model_validation}.
The model predicts the relationship between the estimation accuracy and the accumulated runtime of the estimation.
The main observation is that a majority of the data points lie within the runtime model bounds.
These data points cover a wide range of layer fidelities, accuracies, and expectation values.
Accordingly, we conclude that the runtime model is sufficiently accurate to be used in coarse resource estimations, such as those generated in the main text.

We discuss some of the observed discrepancies between the model and the data. 
In some of the settings (layer fidelity and expectation value), the simulated runtime deviates from the bounds of the runtime model for either high- or low-accuracy.
The deviations for the high accuracy (i.e. small $\varepsilon$) regime are simpler to rationalize.
The Bayesian nature of the RAE approach means that an estimate can deviate far from the true value with small probability.
These bad estimates tend not to improve with further high-L samples due to the phenomenon of aliasing, and they can cause the outliers mentioned above. 
In cases where the runtime grows without improvements in accuracy, we expect that there has been aliasing leading to a bad estimate in the data.
Assuming that this is the cause of these deviations, the likelihood of their occurrence can be exponentially suppressed by repeating the estimation procedure.

A possible explanation for the deviations in the model for the low-accuracy (large $\varepsilon$) cases is as follows.
The analytical derivation of the runtime model bounds in \cite{wang2021minimizing} approximates the runtime as growing continuously as the posterior distribution variance decreases.
This approximation will hold better in the large-runtime (high accuracy) regime, but will tend to underestimate the runtime in the small-runtime regime.
The reason we expect the simulated runtime to be higher than the predicted runtime is that we expect the discretization in the simulated setting to lead to suboptimal choice of layer-number L during each step of the inference process.
This suboptimal choice of L would cause the accuracy to not improve as much as in the idealized continuous case of the model.


\section{Quantification of circuit compilation overhead}
\label{sec:compilation}


In this section we describe the methods used to calculate the circuit depths of the operations used in RAE.
The circuit depths are then used to compute the circuit fidelities as follows.
We model the circuit layer fidelity $f$ as the product of the fidelities of all two-qubit gates in the circuit.
The effective two-qubit gate count is modeled as the depth times half the number of qubits.
The underlying assumption is that, even if a qubit is not subject to an operation in a given layer of the circuit, the decoherence it experiences effectively translates into a loss of fidelity comparable to an imperfect two-qubit gate.

The RAE circuit is composed of a single ansatz circuit $A$ for the state preparation, followed by $L$ layers of RAE iterates $U=AR_0A^{\dagger}P$.
The Pauli gates of $P$ are single-qubit operations so we ignore their contribution to the circuit depth and thus their contribution to reducing the circuit fidelity.
Furthermore, in the accounting of fault-tolerant resources, these Pauli gates contribute negligible error and negligible time relative to other gates because they are Clifford gates.
Therefore, the important accounting we need is of the circuit depths for the ansatz and phase flip operations $D_A$ and $D_R$, respectively. We determine these quantities for various compilation strategies that arise from the device's connectivity of the qubits. These quantities are then used to determine the required logical gate error rate $\overline{r}_g$ and the runtime of each layer $\tau_l$.
We assess the circuit costs associated to the different compilation strategies enabled by the all-to-all connectivity of ion trap devices and the limited 2D connectivity of superconducting qubit devices.

\paragraph{Ansatz circuit}
The ansatz circuit we use in our analysis is the hardware-efficient ansatz (HEA) described in Section III D of \cite{gonthier2020identifying}.
In the case of a linear connectivity, a single layer of this ansatz comprises two layers of nearest-neighbor two-qubit gates. 
We make the optimistic assumption that the number of HEA layers needed to prepare a sufficiently accurate approximation to the ground state is twice the number of qubits (or spin orbitals).
In the case of a two-dimensional connectivity, we will assume that the ansatz is implemented in a ``snake-like'' arrangement in the order of the Jordan-Wigner encoding of the spin-orbitals.
In the case of all-to-all connectivity we will assume that the device enables an ansatz compilation strategy that affords a reduction in gate count and depth by a factor of two.
We believe this to be a conservative assumption because the compilation from a circuit with all-to-all connectivity into a circuit with 2D connectivity typically incurs an overhead cost that grows with the number of qubits.


\paragraph{Phase flip operation}
In our analysis, the most substantial variability in the compilation is in the phase flip operation $R_0=\mathbb{I} -2\ket{0}\! \bra{0}^{\otimes N}$.
Devices with more connectivity will allow for a compilation of the phase flip operation which uses fewer elementary gates and has shorter depth.

The connectivity of the qubits is determined by the elementary 
multi-qubit gates the device can implement, which, in some cases, is determined by the physical layout of the qubits.
We consider two connectivity-dependent compilations of the phase-flip operation as given in \cite{holmes2020impact}.
We name the compilations according to their assumed connectivity:
\begin{itemize}
    \item Two-dimensional: a planar array of qubits connected in a square grid, similar to the qubit layout of some superconducting qubit architectures \cite{ai2021exponential}.
    \item All-to-all: any pair of qubits can be coupled via an elementary gate, which is similar to gates on ion trap devices \cite{debnath2016demonstration}.
\end{itemize}
The characteristics of these various compilation strategies are summarized in Table \ref{tab:compilation}.

\begin{table}
\footnotesize
\centering
\begin{tabular}{|l|l|l|l|l|l|l|}
\hline
 Connectivity & \multicolumn{2}{l|}{Two-dimensional} & \multicolumn{2}{l|}{All-to-all} \\ \hline
 Circuit component &    Ansatz   &  Phase flip     &       Ansatz   &  Phase flip      \\ \hline
\makecell{Two-qubit circuit \\depth} &   N    &   192(N-3)(N-1)    &    N/2     &    32N-96   \\ \hline
\makecell{Effective number of \\gates} &   N$^2$/2    &     96(N-3)(N-1)N   &    N$^2$/4  &    32(N-3)N/2     \\ \hline
\end{tabular}
\centering
\caption{
Summary of compilation circuit costs. Since we are interested in the decay of coherence due to quantum operations (including the idle operation), we will compute the two-qubit circuit depth and then multiply by $N/2$ to get the effective number of two-qubit gates.}
\label{tab:compilation}
\end{table}


\section{Allocation of samples over Pauli terms}
\label{sec:shot_allocation}
We now describe the method of allocating samples over the different Pauli expectation value estimates. The Hamiltonian of interest is decomposed into a linear combination of Pauli terms
\begin{align}
    \sum_i \mu_i P_i,
\end{align}
with coefficients $\mu_i$ and Pauli observables $P_i$.
The objective is to estimate the expected energy of a quantum state $\ket{A}=A\ket{0^N}$.
The true expectation value is a linear combination of Pauli expectation values
\begin{align}
    \bra{A}H\ket{A}=\sum_i \mu_i \bra{A}P_i\ket{A}=\sum_i \mu_i \Pi_i
\end{align}
where we have introduced $\Pi_i=\bra{A}P_i\ket{A}$.
The estimation strategy estimates the expectation value of each Pauli operator $\hat{\Pi}_i$ separately, giving the energy estimate $\hat{E}$ as
\begin{align}
    \hat{E}=\sum_i  \mu_i\hat{\Pi}_i.
\end{align}
For a fixed unbiased estimation strategy used to obtain the $\hat{\Pi}_i$, the total runtime $T_i$ and mean squared error $\varepsilon^2_i$ of each estimator determine the total runtime $T$ and total mean squared error $\varepsilon^2$ of the estimator $\hat{E}$,
\begin{align}
    T &= \sum_i T_i\\
    \varepsilon^2 &= \sum_i \mu_i^2 \varepsilon^2_i.
\end{align}
We fix the target mean squared error (MSE) at chemical accuracy $\varepsilon^2=\bar{\varepsilon}^2$ and aim to determine the estimation strategy for each $\hat{\Pi}_i$ which minimizes the total runtime $T$.

Using the single-term estimation runtime model validated in Appendix \ref{sec:single-term-runtime}, for each Pauli expectation estimation, we model the runtime $T_i$ to target MSE $\varepsilon_i^2$ as
\begin{align}
    T_i = \frac{\omega}{2}\left(\frac{\lambda}{\varepsilon_i^2}+\frac{1}{\sqrt{2}\varepsilon_i}+\sqrt{\left(\frac{\lambda}{\varepsilon_i^2}\right)^2+\left(\frac{\sqrt{8}}{\varepsilon_i}\right)^2}\right).
\end{align}

With the above relationships established, we determine the optimal allocation of runtime using the following numerical optimization:
\begin{align}
    &\min_{\bar{\varepsilon}^2=\sum_i \mu_i^2\varepsilon^2_i}\sum_i T_i\nonumber \\
    &=\min_{\bar{\varepsilon}^2=\sum_i \mu_i^2\varepsilon^2_i}\sum_i \frac{\omega}{2}\left(\frac{\lambda}{\varepsilon_i^2}+\frac{1}{\sqrt{2}\varepsilon_i}+\sqrt{\left(\frac{\lambda}{\varepsilon_i^2}\right)^2+\left(\frac{\sqrt{8}}{\varepsilon_i}\right)^2}\right),
\end{align}
where $\omega$ is proportional to the duration of each layer in the RAE circuit (c.f. Eq. 77 in \cite{wang2021minimizing}), $\mu_i$ are the coefficients in the Pauli decomposition of the Hamiltonian, and $\lambda$ is the fidelity decay parameter in the noise model of the RAE likelihood function.

We introduce a Lagrange multiplier $\Lambda$ to incorporate the constraint and solve for the extreme point
\begin{align}
    0=&\frac{d}{d\varepsilon_i}\left(\sum_i T_i+\Lambda\left(\sum_i \mu_i^2\varepsilon^2_i-\bar{\varepsilon}^2\right)\right)\\
    =&\frac{\omega}{2}\left(\frac{-2\lambda}{\varepsilon_i^3}+\frac{-1}{\sqrt{2}\varepsilon_i^2}+\frac{d}{d\varepsilon_i}\sqrt{\left(\frac{\lambda}{\varepsilon_i^2}\right)^2+\left(\frac{\sqrt{8}}{\varepsilon_i}\right)^2}\right)\\
    &+\Lambda\mu_i^2\varepsilon_i
    \nonumber
\end{align}
In order to arrive at an analytic solution for the $\varepsilon_i$, we approximate the hypotenuse expression above as simply the sum of the two legs of the hypotenuse, which upper bounds the contribution to the runtime via the triangle inequality.
The consequence of this approximation is that the allotment of runtime to each term will be suboptimal leading to an increased overall runtime (relative to that of the optimal allotment).
\begin{align}
    0&=\frac{d}{d\varepsilon_i}\left(\sum_i T_i+\Lambda\left(\sum_i \mu_i^2\varepsilon^2_i-\bar{\varepsilon}^2\right)\right)\\
    &\approx\frac{\omega}{2}\left(\frac{-2\lambda}{\varepsilon_i^3}+\frac{-1}{\sqrt{2}\varepsilon_i^2}+\frac{-2\lambda}{\varepsilon_i^3}+\frac{-\sqrt{8}}{\varepsilon_i^2}\right)+\Lambda\mu_i^2\varepsilon_i\\
    &=\frac{-2\omega\lambda}{\varepsilon_i^3}+\frac{-\alpha}{\varepsilon_i^2}+\Lambda\mu_i^2\varepsilon_i,
\end{align}
where $\alpha=\frac{1}{2}(\sqrt{1/2}+\sqrt{8})$.
The MSEs are determined by solutions to
\begin{align}
    -2\omega\lambda-\alpha\varepsilon_i+\Lambda\mu_i^2\varepsilon_i^4=0.
\end{align}
Letting $a=\alpha/\Lambda\mu_i^2$ and $b=2\omega\lambda/\Lambda\mu_i^2$, we must solve the quartic equation $\varepsilon^4_i=a\varepsilon_i+b$.
The solutions in the shot-noise limit and Heisenberg limit extremes correspond to $a\approx 0$, giving $\varepsilon_i^2=b^{1/2}$, and $b\approx0$ giving $\varepsilon_i^2=a^{2/3}$, respectively.

We approximately solve the quartic to obtain $\varepsilon_i^2\approx b^{1/2}+a^{2/3}$.
Plugging back in the relevant values gives
\begin{align}
    \varepsilon_i^2 = \frac{\sqrt{2\omega\lambda}}{\Lambda^{1/2}|\mu_i|}+\frac{\alpha^{2/3}}{\Lambda^{2/3}|\mu_i|^{4/3}}
\end{align}
To obtain $\Lambda$ we plug the above expression into the constraint equation,
\begin{align}
    \bar{\varepsilon}^2=\frac{\sqrt{2\omega\lambda}}{\Lambda^{1/2}}\sum_i|\mu_i|+\frac{\alpha^{2/3}}{\Lambda^{2/3}}\sum_i|\mu_i|^{2/3}.
\end{align}
Arranging into a polynomial in $\Lambda^{1/6}$, we obtain
\begin{align}
    (\Lambda^{1/6})^4\bar{\varepsilon}^2=(\Lambda^{1/6})\sqrt{2\omega\lambda}\sum_i|\mu_i|+\alpha^{2/3}\sum_i|\mu_i|^{2/3}.
\end{align}
We will solve this quartic polynomial numerically to ensure that the normalization is correct and that the chemical accuracy constraint is satisfied.
Let $\Lambda_*$ be the numerical solution.
The total runtime $T_*$ is then determined by the following procedure.
First, we solve for $\Lambda$ numerically.
Then we evaluate each MSE according to $\varepsilon_i^2=\frac{\sqrt{2\omega\lambda}}{\Lambda^{1/2}|\mu_i|}+\frac{\alpha^{2/3}}{\Lambda^{2/3}|\mu_i|^{4/3}}$.
Finally we evaluate the overall runtime as
\begin{align}
    T_* = \sum_i \frac{\omega}{2}\left(\frac{\lambda}{\varepsilon_i^2}+\frac{1}{\sqrt{2}\varepsilon_i}+\sqrt{\left(\frac{\lambda}{\varepsilon_i^2}\right)^2+\left(\frac{\sqrt{8}}{\varepsilon_i}\right)^2}\right).
\end{align}

\section{Model of fault-tolerant quantum computation overhead}
\label{sec:fault_tolerance}
In order to sufficiently reduce the estimation runtimes, we must sufficiently reduce the error rates of the quantum operations.
The scalable approach to reducing such error rates is achieved with quantum error correction.
While quantum error correction suppresses the error rates of gates and measurements, it incurs an additional cost in terms of physical qubits and processing time.
We will analyze these costs of quantum error correction assuming a state-of-the-art implementation of the surface code \cite{wang2009threshold}.
Following the analysis in \cite{huggins2020virtual}, we consider a quantum computer architecture which runs the surface code with physical gate error rates of $10^{-3}$.

By increasing the code distance $d$ the logical error rates are reduced as
\begin{align}
    \varepsilon = 10^{-(d+3)/2}.
\end{align}
To protect each operation with a distance-$d$ code requires $N=2d^2$ physical qubits.
Following the assumptions of \cite{huggins2020virtual} (that the synthesis of each gate will require, on average $100d$ surface code cycles), then the fidelity of each gate will be
\begin{align}
    f=(1-10^{-(d+3)/2})^{100d}\approx 1 - d10^{-(d-1)/2}.
\end{align}
The gates which are enumerated in this model include arbitrary-angle single-qubit rotations as well as all two-qubit gates.
Optimistic estimates \cite{wang2009threshold} give surface cycle times of $1\mu$s.
Thus, the time needed to implement the above logical gate is $100d \mu$s.

We incorporate these quantum error correction overheads into our runtime estimates by assuming that each single layer of logical gates requires a runtime of $100d\mu$s. 
Then, both the gate layer runtime and logical gate fidelity are determined by the code distance $d$.

%

\end{document}